\documentclass{PoS}

\title{Towards a Direct Measurement of the Quark Orbital Angular Momentum Distribution}

\ShortTitle{Quark Orbital Angular Momentum}

\author{\speaker{Simonetta Liuti}\thanks{U.S. D.O.E. grants DE-FG02-01ER4120(S.L.), DE-FG02-96ER40965 (M.E.)}\\
        University of Virginia and INFN, Laboratori Nazionali di Frascati \\
        E-mail: \email{sl4y@virginia.edu}}

\author{Aurore Courtoy  \\
        \\ Universidad de Guanajuato, Mexico and Universit\'e de Li\`ege, Belgium \\
        E-mail: \email{aurore.courtoy@ulg.be.ac}}
        
\author{Michael Engelhardt\\
        New Mexico State University\\
        E-mail: \email{engel@nmsu.edu}}

\author{Abha Rajan \\
        University of Virginia\\
        E-mail: \email{ar5xc@virginia.edu}}

\abstract{We discuss two different definitions of partonic orbital angular momentum given in the literature in terms of the Fourier transform of a Wigner distribution, $F_{14}$,  and a twist three generalized parton distribution, $\tilde{E}_{2T}$, respectively. 
We derive a relation between the two definitions which reflects their specific spin and intrinsic transverse momentum/transverse space correlations as well as their gauge link structure.}

\FullConference{XXIII International Workshop on Deep-Inelastic Scattering\\
		27 April - May 1 2015\\
		Dallas, Texas}

\begin{document}

\section{Introduction}
Partonic Orbital Angular Momentum (OAM) is generated inside the proton as a consequence of the quark and gluon transverse motion about the system's center of momentum. OAM was identified with the missing component in the proton's spin sum rule \cite{Jaffe:1989jz}.  Many efforts have since then been directed at exploring the issues behind the need to ensure  a gauge invariant decomposition,  within QCD , of angular momentum  into its spin and orbital components (see {\it e.g.}  \cite{Leader:2013jra} for a review). These efforts have borne important consequences. They brought on one side Ji to formulate his sum rule where the total angular momentum is expressed in terms of new observables stemming from Generalized Parton Distributions (GPDs) \cite{Ji:1996ek}. On the other hand, a modified version 
of the original sum rule in \cite{Jaffe:1989jz} was proposed where each term observes gauge invariance \cite{Bashinsky:1998if}. The two definitions, Jaffe and Manohar's (JM)  \cite{Jaffe:1989jz}, and Ji's (Ji) \cite{Ji:1996ek}, have been shown to be connected: JM's includes final state interactions of the struck quark with the proton's remnant, and it reduces to Ji's when these are disregarded \cite{Burkardt:2012sd}.  
Notwithstanding these important developments the problem of the observability of OAM has remained an unsolved question for years as clearly stated in Ref.\cite{Jaffe:1999ze}: ``.. it appears that the distributions $L_q(x,Q^2)$ and $L_g(x,Q^2)$ are not experimentally accessible. So the value of the sum rule is obscure. "

The quark OAM has been written using Generalized Transverse Momentum Distributions (GTMDs) \cite{Lorce:2011kd,Lorce:2011ni,Hatta:2011ku}. GTMDs can be considered a hybrid of the GPDs and of the Transverse Momentum Distributions (TMDs), which in turn appear as two different limiting cases of a GTMD. 
GTMDs are directly connected through Fourier transformation to the quark Wigner distributions --  the quantum mechanical version of phase space distributions.
The importance of writing OAM in terms of GTMDs is that they provide a formal framework to perform calculations in both lattice QCD \cite{Engelhardt:2014eea} and in phenomenological models
\cite{Lorce:2011kd,Burkardt:2008ua,Courtoy:2013oaa,Mukherjee:2014nya}. Nonetheless, the question of observability is still posing issues. As of yet
``...it is not known how to extract Wigner distributions or GTMDs from experiments" \cite{Lorce:2011kd}.

An alternative way to obtain quark OAM was given by Polyakov and collaborators 
\cite{Penttinen:2000dg,Kiptily:2002nx} (see also \cite{Hatta:2012cs}), who using OPE techniques showed that a specific twist-three GPD, $\tilde{E}_{2T}$, written in the Wandzura Wilczek approximation 
\cite{Wandzura:1977qf}, 
corresponds to a form containing the twist-two GPDs, $H$ and $E$, defining the total angular momentum, $J_q$, and the spin term, $\Delta \Sigma$, which is known and measurable in polarized DIS. Specifically, by taking the second moment of this form one fulfills Ji's sum rule  \cite{Ji:1996ek}.
The advantage of this definition is that  it allows us to verify all three terms in the sum rule for the quark sector, since GPDs are readily measurable.  Indeed, investigations of  experimental hard scattering processes/observables that measure OAM directly have already started. 

The GTMD and the twist-three GPD based definitions of OAM have seemed so far to be unrelated despite purporting to describe the same physical quantity. 
Notice that the two different definitions refer to different types of parton distributions, regardless of the gauge link structure. The latter was studied using the GTMD/Wigner distribution framework in Ref.\cite{Burkardt:2012sd}. In this situation it is important to find the connection between the two.

This contribution is dedicated to a preliminary study of the issue of connecting definitions of OAM in QCD. More details can be found in a forthcoming paper on the subject. 

\section{Quark Orbital Angular Momentum}
OAM enters two gauge invariant decompositions of the proton's angular momentum in terms of partonic degrees of freedom. The one derived by Ji  \cite{Ji:1996ek} is,  
\begin{equation}
\label{Ji:eq}
\frac{1}{2} = S_q + L^{\rm{Ji} }_q + J^{\rm{Ji} }_g = J^{\rm{Ji} }_q + J^{\rm{Ji} }_g  \ ,
\end{equation}
where $S_q$, the total (summed over flavor) quark spin is given by the integral of the helicity distribution, $g_1(x)$, $L^{\rm{Ji} }_q$ is the quark OAM, and $J^{\rm{Ji} }_{q(g)}$ is the quark (gluon) angular momentum given by the second moment of the unpolarized Generalized Parton Distribution (GPD) $H$ plus the spin flip GPD, $E$.
The decomposition by Jaffe and Manohar (JM) reads \cite{Jaffe:1989jz}, 
\begin{equation}
\label{JM:eq}
\frac{1}{2} = S_q + L^{\rm{JM} }_q + S^{\rm{JM} }_g + L^{\rm{JM} }_g.
\end{equation}
The only term  both definitions have in common is the
quark spin contribution $S_q $, while
$L^{\rm{Ji} }_q \neq L^{\rm{JM} }_q $, and $J^{\rm{Ji} }_g \neq S^{\rm{JM} }_g + L^{\rm{JM} }_g$; $S^{\rm{JM} }_g$, the gluon spin, can also be measured in experiments via the gluon helicity structure function, $\Delta G$. 

Quark OAM can be defined as,
\begin{eqnarray}
\label{Lu}
L^{\cal U}_q =\int dx \int d^2 {\bf k}_T \int d^2{\bf b} \,  ({\bf b} \times \bar{\bf k}_T)_3 {\cal W}^{\, \cal U}(x,\bar{\bf k}_T,{\bf b}),
\end{eqnarray}
where ${\cal W}^{\, \cal U}$ is a Wigner distribution given by the Fourier transform of the quark-quark off-forward unintegrated correlator as,
\begin{eqnarray}
\label{Wigner}
 {\cal W}^{\, \cal U}(x,\bar{\bf k}_T,{\bf b}) & = &\int  \frac{d^2 \Delta_T}{(2 \pi)^2} \, e^{i {\bf b} \cdot  \Delta_T}  \left[ W^{\gamma^+}_{+ +} - W^{\gamma^+}_{- -} \right]. 
 \end{eqnarray}
We have taken the same proton helicities in the initial and final states,  $\Lambda = \Lambda'$, of the correlator parametrized as in  \cite{Meissner:2009ww},  
\begin{eqnarray}
\label{Wu}
&&W_{\Lambda \Lambda}^{\gamma^+}  =  \displaystyle\int \frac{d z^- \, d^2{\bf z}_T}{(2 \pi)^3} e^{ixP^+ z^- - i \bar{\bf k}_T\cdot {\bf z}_T}  \langle p', \Lambda  \mid 
 \bar{\psi} (-z/2) \gamma^+  {\cal U}(-z/2,z/2)  \psi\left(z/2 \right)  \mid p, \Lambda \rangle \left.  \right|_{z^+=0}, \nonumber \\
 & = & \frac{1}{2P^+} \overline{U}(p',\Lambda) \left(\gamma^+ F_{11} + \frac{i \sigma^{ij} \bar{k}_i\Delta_j}{M^2} F_{14} \right) U(p,\Lambda).
\end{eqnarray}
%
In Eqs.(\ref{Lu},\ref{Wigner},\ref{Wu}), $p=P+\Delta/2$, $p'=P-\Delta/2$, $P= (p+p')/2$, $k=\bar{k}+\Delta/2$, $k'=\bar{k}-\Delta/2$, $\bar{k}= (k +k')/2$, and the skewness parameter, $\xi=\Delta^+/P^+ =0$, hence $t=\Delta^2 \equiv - {\Delta}_T^2$.
$F_{11}$ and $F_{14}$ are GTMDs describing an unpolarized quark inside an unpolarized proton, and an unpolarized quark inside a longitudinally polarized proton, respectively \cite{Meissner:2009ww}.
\footnote{We follow the notation of Ref.\cite{Meissner:2009ww}, where $F_{1n}, G_{1n} (n=1,4)$, indicate twist-two GTMDs, and $F_{2n}, G_{2n} (n=1,8)$ twist-three GTMDs, respectively in the vector ($F$) and axial-vector ($G$) sectors.}
One can then relate $L^{\cal U}_q$  to the $k_T$ moment of $F_{14}$ through 2D Fourier transformation \cite{Hatta:2011ku,Lorce:2011ni,Meissner:2009ww},
\begin{eqnarray}
\label{F14_1}
L^{\cal U}_q = F_{14}^{(1)} & \equiv & \int d^2 \bar{\bf k}_T \, \frac{\bar{k}_T^2}{M^2} F_{14}(x, 0, \bar{k}_T^2, \bar{k}_T \cdot \Delta_T, \Delta_T^2). 
\end{eqnarray}
The superscript  ${\cal U}= $(JM, Ji), in Eq.(\ref{Lu}) specifies the gauge link which is different in the two descriptions \cite{Burkardt:2012sd,Hatta:2011ku}: for JM, one has  a staple-shaped gauge link taking into
account the final state interactions experienced by the struck quark. By contrast, Ji's OAM is characterized by  a straight gauge link, {\it i.e.} by the absence of final state interactions.   
As is familiar from TMD studies the two terms can be related as,
\begin{eqnarray}
\label{Burkardt:eq1}
L^{\rm{JM} }_q  =  L^{\rm{Ji} }_q  + \langle T_3 \rangle
\end{eqnarray}
where
\begin{eqnarray} 
\label{Burkardt:eq2}
 \langle T_3 \rangle  =  -g \left( x G^{+2} - y G^{+1} \right)
 =   \!\! \int d z^-  d^2{\bf z}_T  \langle p, \Lambda \!\! \mid 
 \bar{\psi} ({\bf z}) \gamma^+  \!\!\! \int_{z^-}^\infty  \!\! d z'^- T_3 (z'^-,{\bf z}_T) \! \psi({\bf z})  \!\! \mid p, \Lambda \rangle \nonumber \\
\end{eqnarray}
is an off-forward extension of a Qiu-Sterman term \cite{Qiu:1991pp}. This appears to have the physical meaning of a torque (a final state interaction) exerted on the outgoing quark by the color-magnetic field produced by the spectators\cite{Burkardt:2012sd}. 

%
Another way of describing quark OAM was obtained by Polyakov and collaborators \cite{Penttinen:2000dg,Kiptily:2002nx}, extending  the derivation of  the WW relations to the GPD sector.  Using OPE techniques it was shown that OAM can be described through a twist-three GPD, $\tilde{E}_{2T}$, appearing in the parametrization of the vector sector (we implement a newer, uniform notation introduced in Ref.\cite{Meissner:2009ww}),
\begin{eqnarray}
\label{tw3_metz}
W^{\gamma^i}_{\Lambda' \Lambda} & = & \frac{M}{P^+} \frac{1}{2 P^+}  \overline{U}(p',\Lambda')  \left[%
 i \sigma^{+i} H_{2T}(x,\xi,t) +
 \frac{\gamma^+ \Delta^i - \Delta^+ \gamma^i}{2M} E_{2T}(x,\xi,t)   \right. \nonumber \\
& + & \left.  \frac{P^+ \Delta^i - \Delta^+ P^i}{M^2}  \widetilde{H}_{2T}(x,\xi,t)  +
\frac{\gamma^+ P^i - P^+ \gamma^i}{M} \widetilde{E}_{2T}(x,\xi,t) \right]    U(p,\Lambda), 
 \end{eqnarray} 
The WW part of the twist-three GPD, $\tilde{E}_{2T}$, is given by the following combination of twist-two GPDs,
\begin{eqnarray}
\label{polyakov}
\tilde{E}_{2T}^{WW}(x,\Delta_T) = - \int\limits_{x}^{1}  \frac{dy}{y} (H+ E) +   \int \limits_{x}^{1} \frac{dy}{y^2} \tilde{H}.
\end{eqnarray}
The negative of the second moment of $\tilde{E}_{2T}^{WW}$ coincides with the quark OAM from the Ji sum rule,  namely,
\begin{eqnarray}
 L_q^{\rm Ji} & = &J_q^{\rm Ji}-S_q \equiv J_q^{\rm Ji}- \frac{1}{2} \Delta \Sigma   \Rightarrow  \nonumber \\
-  \int\limits_{-1}^1  dx x \, \tilde{E}_{2T}^{WW} & = & \frac{1}{2} \int\limits_{-1}^1 dx x (H+ E) - \frac{1}{2}  \int \limits_{-1}^1 dx \, \tilde{H}.
\end{eqnarray} 
Only a straight gauge link structure applies here, owing to the fact that no (generalized) transverse momentum dependent distribution is directly involved.
%
Notice that this derivation does not make use of GTMDs to describe OAM. 
This leads to an interesting situation where even if one remains in the context of Ji's decomposition of angular momentum, OAM can be described simultaneously by two different structure functions, a GTMD on one side, and a twist-three GPD on the other.

\section{Relation between the two definitions of OAM}
We show that a relation can be found between the two descriptions in terms of  $F_{14}$,  and  $\tilde{E}_{2T}$, respectively, that reflects both the underlying spin correlation and the gauge link structures of partonic OAM. The relation reads,
%
\begin{eqnarray}
\label{relation1}
 x \tilde{E}_{2T} =  \int d^2 \bar{k}_T  \, \frac{\bar{k}_T^2}{M^2} \, F_{14}  - 2 \int d^2 \bar{k}_T  \,  \left[ \frac{\bar{k}_T \cdot {\Delta}_T}{\Delta_T^2}  \, F_{12} + F_{13} \right]
 + \int d^2 \bar{k}_T  \, \frac{\bar{k}_T \cdot {\Delta}_T}{\Delta_T^2}  \, G_{14}  + G^{tw 3},
\end{eqnarray}
All quantities in Eq.(\ref{relation1}) are  evaluated at $(x,0,0)$ {\it i.e.} in the forward limit. $\tilde{E}_{2T}$ can be written in terms of GTMDs as  \cite{Meissner:2009ww},
\begin{eqnarray}
\label{G2}
\tilde{E}_{2T}(x,0,t) & = & -2 \int d^2 \bar{k}_T \left[ \frac{\bar{k}_T \cdot \Delta_T}{\Delta_T^2} F_{27}(x, 0, \bar{k}_T^2,k\cdot \Delta,\Delta_T^2) +  F_{28}(x, 0, \bar{k}_T^2,k\cdot \Delta,\Delta_T^2) \right ] 
\end{eqnarray}
The terms involving $G_{14}$, $F_{12}$, and $F_{13}$,  integrate to the helicity GPD, $\tilde{H}$, and to $H+E$, respectively,
\begin{equation}
\tilde{H}   =  \int d^2 \bar{k}_T   G_{14} \;\;\;\;\;\; H+E   =  2 \int d^2 \bar{k}_T  \, \left[ \frac{\bar{k}_T \cdot {\Delta}_T}{\Delta_T^2}  \, F_{12} + F_{13} \right]
\end{equation}
$G^{tw 3}$, is a genuine twist three term to be discussed later. As shown in detail in a forthcoming publication, by implementing the additional relation,
\begin{equation}
\label{relation2}
\frac{d}{dx}  \int d^2 \bar{k}_T  \, \frac{\bar{k}_T^2}{M^2} \, F_{14} = \tilde{E}_{2T} \,\, \Rightarrow \,\, \int dx  \int d^2 \bar{k}_T  \, \frac{\bar{k}_T^2}{M^2} \, F_{14} = \int dx \, x \tilde{E}_{2T}, 
\end{equation}
derived by observing that $F_{14}$, and $\tilde{E}_{2T}$ admit a common substructure in terms of covariant amplitudes \cite{Meissner:2009ww,Mulders:1995dh}, one can eliminate $F_{14}$ from Eq.(\ref{relation1}), thus obtaining Eq.(\ref{polyakov}). 

The procedure used to derive Eq.(\ref{relation1}) is obtained following the approach used  by Mulders and collaborators to derive WW type relations for the twist-three distributions $g_T$ and $h_L$ \cite{Mulders:1995dh},  extending them off-forward. 
The basis of our derivation is therefore the same as in Ref.\cite{Mulders:1995dh} in that we use directly the {\em nonlocal} quark-quark and quark-gluon-quark correlators. This gives a more transparent interpretation than from using the standard methods of OPE, by emphasizing the role of partonic transverse momentum and off-shellness, whereby OAM is defined starting from nonlocal, $\bar{k}_T$ unintegrated, off-forward matrix elements.

It is important to mark the distinction between {\it geometric/canonical twist} and {\it dynamical twist}. Canonical  twist, $\tau$, is defined formally as the canonical dimension minus the Lorentz spin of the local operators that enter the expansion of the various observables/currents in inverse powers of $Q^2$ (OPE). 
Dynamical twist, $t$,  is defined by projecting out the quark field's good, $\gamma^- \gamma^+ \psi$, and bad, $\gamma^+ \gamma^- \psi$, components, respectively. 
The order of dynamical and canonical twist does not match beyond order two: 
contributions  with the same power in $M/Q$, or same dynamical twist, can be written in terms of matrix elements of operators with different canonical twist, $\tau$. 
The WW relations between matrix elements of operators of different dynamical and same canonical twist encode this mismatch.  

Eq.(\ref{relation1}) represents an extension of this physical description to the off-forward case.  

\section{Conclusions}
The GTMD \cite{Lorce:2011kd,Lorce:2011ni,Hatta:2011ku} and twist-three GPD \cite{Penttinen:2000dg,Kiptily:2002nx,Hatta:2012cs} based definitions of quark OAM have different origins. The GTMD one is obtained directly from the matrix element of the angular momentum tensor in QCD, where the spin and orbital parts are evaluated in the framework of either JM's \cite{Jaffe:1989jz} or Ji's \cite{Ji:1996ek} decompositions of angular momentum.  The twist-three GPD definition, stemming from Ji's decomposition, results from an off-forward extension of the Wandzura Wilczek relations originally derived for the twist three PDF, $g_2$ \cite{Jaffe:1996zw}.

We presented an initial study of the connection between these two seemingly unrelated definitions of OAM. Our result is contained in Eqs.(\ref{relation1}) and (\ref{relation2}). From these relations we conclude that both definitions represent directly OAM.
Our derivation applies strictly to Ji's definition of OAM. A second step including final state interactions \cite{Burkardt:2012sd} is then needed to obtain JM's OAM. 

The OAM relations in Eqs.(\ref{relation1}) and (\ref{relation2}) provide a significant link between two types of distributions that allows us 
to connect ongoing calculations of GTMDs in lattice QCD with experimental observables.  A recent study of the helicity structure of  the OAM GPD \cite{Courtoy:2013oaa}, has in fact shown that the GPD $\tilde{E}_{2T}$ can be extracted from the $\sin 2 \phi$ modulation of the Deeply Virtual Compton Scattering longitudinal target spin asymmetry, $A_{UL}$. 

These relations also provide a starting point to explore many open interesting questions concerning possible constraints on the transverse momentum dependent perturbative evolution of $F_{14}$, on one side, and, on the other, the connection with the axial-vector counterpart of $\tilde{E}_{2T}$, given by the transverse spin structure function, $g_2$ \cite{Jaffe:1996zw,Burkardt:2009rf}.  

\vspace{0.5cm}
\noindent
We thank Matthias Burkardt for many useful discussions on this subject, and Harut Avakian and Silvia Pisano for discussions 
on the experimental extraction of twist three GPDs from DVCS data.


\end{document}